# Parametric Resonance and RF-to-THz Frequency Conversion in Semiconductor Plasmonic Crystals

G. R. Aizin[1, 2,*], J. Mikalopas[1], M. Shur[3,4,†]

[1]*Kingsborough College, The City University of New York, Brooklyn, NY 11235, USA*
[2]*The Graduate School and University Center, The City University of New York, New York, NY 10016, USA*
[3]*Rensselaer Polytechnic Institute, Troy, NY 12180, USA*
[4]*Electronics of the Future, Inc., Vienna, VA 22181, USA*

## Abstract

We show that plasma excitations in nanoscale field-effect transistor structures with periodic alternation of gated and ungated regions (plasmonic crystals) differ fundamentally from conventional plasmons in isolated gated or ungated regions. In contrast to the linear dispersion of purely gated plasmons and the square-root dispersion of ungated plasmons, these collective modes also exhibit a parabolic dispersion law characterized by a finite plasmonic effective mass. We call these excitations "rotonic plasmons" emphasizing the analogy to roton-like excitations. The dynamics of rotonic plasmons are governed by a generalized Mathieu equation, describing either resonant or non-resonant parametric excitations of rotonic plasmons depending on damping. These nonlinear resonances can be efficiently driven by gate-voltage pumping, avoiding the spatial nonuniformities and electron drift velocity saturation effects associated with current-driven excitation. Gate-voltage pumping enables much higher terahertz (THz) power levels in plasmonic crystals. More importantly, in contrast to source-drain excitation, gate voltage pumping has the same gate voltage swing over large area transistors or transistor arrays. We develop a unified theory of rotonic plasmons and demonstrate their application for RF to THz frequency multiplication and THz generation. Starting from the general dispersion relation in plasmonic crystals based on coupled gated-ungated regions with two-dimensional electron gas, we derive the parabolic ("rotonic") plasmon spectrum and establish its analogy with roton-like excitations. The analysis predicts parametric instabilities in III–N and III–V plasmonic crystals under gate-voltage pumping. The results confirm that these systems can function as tunable, compact THz sources and detectors suitable for emerging 6G communications and sensing applications.

* gaizin@kbcc.cuny.edu
† shurm@rpi.edu

## I. INTRODUCTION

The exponential growth of global data traffic necessitates a paradigm shift toward sixth-generation (6G) communication systems operating in the sub-terahertz (sub-THz, 100–300 GHz) frequency range [1,2]. Exploiting these higher frequencies promises orders-of-magnitude improvements in data throughput and latency, but presents major challenges in efficient signal generation, detection using existing and emerging semiconductor technologies in the THz and sub-THz regions. Exciting and rectifying electron (or hole) plasma oscillations (plasmons) in periodic grating gate field-effect transistor arrays (plasmonic crystals [3-11]) is one of the most promising approaches to THz signal detection and generation. The frequencies of these plasmons lie in the THz range and tunable by a gate voltage. In plasmonic crystals with a sufficiently long mean free path the coherence of the plasma oscillations excited in the elementary cells is maintained over macroscopic distances. This coherence over large areas and commensurate increase in operating power (proportional to the area of a plasmonic crystal or a plasmonic array) makes plasmonic crystals strong candidates for bridging the "THz gap," offering compact and tunable architectures with potentially high output powers [12-21].

A two-dimensional (2D) electron gas is periodically modulated in grating-gate structures (shown in Fig. 1). Such a modulation could also be achieved by modulating the thickness of the barrier dielectric layer separating the channel and the gate [9], or by using the channels with periodically modulated width [22]. Grating-gate FETs were used in the pioneering experiments resulting in the discovery of the 2D plasmons [23] and theoretical prediction [24] and demonstration [25] of resonant peak splitting in plasmonic frequency response revealing an energy gap in the plasmon energy spectrum.

In this paper, we use the dispersion equation for plasmonic crystals to show the existence of modes that exhibit parabolic dispersion and are characterized by a finite plasmonic effective mass. We call these excitations "rotonic plasmons" emphasizing their analogy to roton-like excitations. The rotonic dispersion law occurs near mode intersections and near plasmonic band edges. We show that the dynamics of rotonic plasmons under gate voltage pumping is governed by a highly nonlinear version of the Mathieu equation [26,27]. The highest nonlinearity is achieved when the gate voltage modulation periodically switches the gated regions from above to below threshold regime and vice versa – with these areas operating as off-to-on switches. This new rotonic regime supports resonant or non-resonant parametric excitations depending on damping. The gate voltage pumping eliminates spatial nonuniformities typical of current-driven excitation - a major problem in conventional plasmonic crystal excitation by channel current. Our unified theory of rotonic plasmons demonstrates their application for RF to THz frequency conversion and THz generation. The analysis predicts parametric instabilities in III–N and III–V plasmonic crystals under gate-voltage modulation pumping. Our results show that these systems can operate as tunable, compact THz sources and detectors suitable for 6G communication and sensing applications.

The paper is organized as follows: Section II presents the derivation and solution of the rotonic plasmon dispersion equation. It is applied for the analysis of plasmonic crystals with large ratio of carrier densities in gated and ungated regions (the "low-high" plasmonic crystals). Section III analyzes parametric excitation in AlGaN/GaN and AlGaAs/GaAs systems using nonlinear Mathieu dynamics. Section IV summarizes the main conclusions.

## II. THEORETICAL MODEL

Plasmonic crystals demonstrate new fundamental effects supported by their unique band energy spectrum. In contrast to conventional plasmons in isolated gated or ungated regions, plasma excitations in periodic plasmonic crystal structures exhibit a fundamentally different dispersion law, as demonstrated in a number of recent experiments [19-21].

Fig. 1 shows a schematic of a one-dimensional plasmonic crystal structure. The grating gate introduces periodic modulation of electron density and gate screening in the 2D electron channel of the FET.

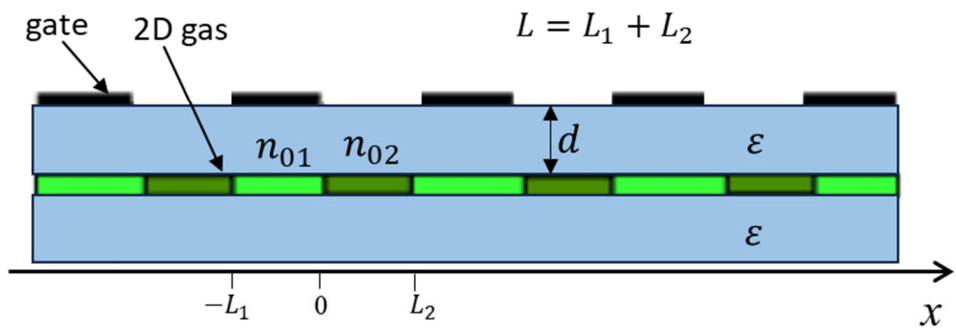

Fig. 1. Schematic of a plasmonic crystal one-dimensional periodic structure. 2D electron channel has gated sections of length $L_1$ with electron density $n_{01}$ and ungated sections of length $L_2$ with electron density $n_{02}$, respectively, $d$ and $\varepsilon$ are the thickness and the electric permittivity of the barrier dielectric layer

The plasmonic crystal band energy spectrum is formed if and only if the plasmon mean free path $L_{pl} = v_{pl}\tau$ exceeds the modulation period $L$, i.e., $v_{pl}\tau \gg L$. Here, $v_{pl}$ is characteristic plasma velocity and $\tau$ is the plasmon relaxation time (which is of the same order as the electron momentum relaxation time). Since the plasma frequency $\omega_{pl}$ in the individual elementary cell is of the order of $v_{pl}/L$, the plasmonic crystal band energy spectrum is formed if $\omega_{pl}\tau \gg 1$. This last inequality is the condition of sustainability of the plasma oscillations in a single FET. However, the coherent multi-gated transistor structures can be orders of magnitude larger than the size of an elementary cell $L$. This situation is analogous to a single crystal, which is many orders of magnitude larger than the electron mean free path. The larger active area of the plasmonic crystal structure gives plasmonic crystals a clear advantage over single TeraFETs for designing THz detectors, amplifiers and emitters based on the plasma oscillations in the 2D electron channels.

The dispersion equation for a one-dimensional plasmonic crystal was derived for both grating-gated FET structures [8] and for fully gated FETs with periodically modulated electron density [9]. This equation was derived using the hydrodynamic model assuming continuity of the plasmonic current and energy flow at the boundaries between elementary crystal cells. Similar equation was also derived with ballistic boundary conditions [28] assuming voltage continuity across the boundary.

In the following we assume that the gate dielectric thickness $d$ is much smaller than the lengths of the gated ($L_1$) and ungated ($L_2$) sections of the channel: $d \ll L_1, L_2$, see Fig. 1. We also assume small plasmon damping rate: $\omega_{pl}\tau \gg 1$. In this limit, the dispersion equation for the plasmon energy bands in this system takes the form [8,28] (see also the Appendix):

$$\cos kL = \cos\frac{\widetilde{\omega}L_1}{L}\cos\frac{\widetilde{\omega}^2\eta aL_2}{L} - \frac{1}{2}\left(\widetilde{\omega}a + \frac{1}{\widetilde{\omega}a}\right)\sin\frac{\widetilde{\omega}L_1}{L}\sin\frac{\widetilde{\omega}^2\eta aL_2}{L} \qquad (1)$$

In this equation, $k$ is the plasmon Bloch wave vector ($-\pi/L \leq k < \pi/L$), $L = L_1 + L_2$ is the grating period, $\widetilde{\omega} = \frac{\omega L}{v_p}$ is the dimensionless plasma frequency, and $v_p$ is the plasma velocity in the gated sections: $v_p = (e^2 n_{01} d/m^* \varepsilon\varepsilon_0)^{1/2}$. Here, $\eta = n_{01}/n_{02}$ is the density modulation factor, $a = \frac{\varepsilon+1}{\varepsilon}\frac{d}{L}$, $n_{01}$ and $n_{02}$ are equilibrium electron densities in the gated and ungated sections of the 2D channel, respectively, $\varepsilon$ is the electric permittivity of the dielectric layer, and $m^*$ is electron effective mass.

In the limit of fully gated ($L_1 \rightarrow L$) and ungated ($L_1 \rightarrow 0$) channel, the solution of Eq. (1) yields the well-known dispersion law for gated and ungated plasmons:

$$\omega_g(q) = \sqrt{\frac{e^2 n_{01} d}{m^*\varepsilon\varepsilon_0}}\, q \qquad \text{(gated)}$$

$$\omega_u(q) = \sqrt{\frac{e^2 n_{02} q}{m^*(\varepsilon+1)\varepsilon_0}} \qquad \text{(ungated)} \qquad (2)$$

where $q$ is the plasma wave vector.

In Fig. 2, we show the plasmonic energy band spectrum in the reduced Brillouin zone scheme found numerically from Eq. (1) at $\eta = 0.1$ and $L_1/L = 0.5$. In this calculation, we used the values of parameters typical for GaN-based structures: $n_{02} = 1 \times 10^{17}$ m$^{-2}$, $m^* = 0.24 \, m_0$, $\varepsilon = 8.9$, $d = 25$ nm, and $L = 0.5 \mu m$. As expected, energy gaps open in the plasmon energy spectrum. Near the bandgap edges at $k = k_0 = 0, \pm \frac{\pi}{L}$ dependence of the plasma frequency $\omega$ on the Bloch wave vector $k$ is parabolic: $\omega - \omega_0 \propto \delta k^2$ where $\delta k = k - k_0 \ll 1/L$ and $\omega_0$ is plasma frequency at the edge of the gap. These results directly follow from Eq. (1) expanded near $(\omega_0, k_0)$ point.

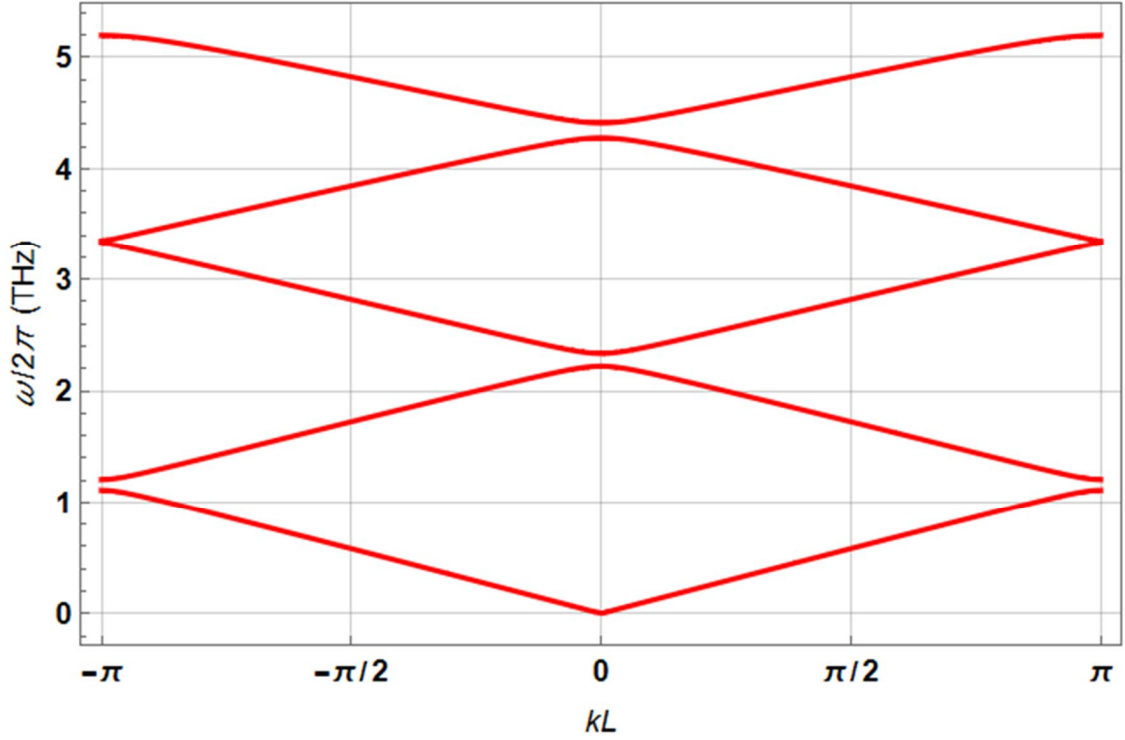

Fig. 2. Energy band plasmonic spectrum in the grating gated 2D electron channel in the reduced Brillouin zone scheme for the density modulated channel with $n_{01}/n_{02} = 0.1$ and the grating filling factor $L_1/L = 0.5$. All other parameters are explained in the text.

As seen from Eq. (1), the plasmon effective mass depends on the system geometry and electron density in the gated regions $n_{01}$. Since the electron density $n_{01}$ can be changed by varying the gate voltage this provides a unique opportunity to control the dynamics of the rotonic plasmons [29].

The plasmonic effective mass can be found from measurements of the plasmonic absorption of an external EM radiation at non-zero angle of incidence when plasmons at $k \neq 0$ are excited. It can also be found by measuring plasma resonances in magnetic field and separating contributions from the cyclotron and plasmonic resonances in the magnetoplasmon spectrum [30].

We estimated the plasmonic effective mass for plasmons propagating in AlGaN/GaN plasmonic crystals using the difference of plasmonic frequencies for adjacent plasma resonant peaks $\Delta\omega$ and parabolic wave vector dependence for $k = 2\pi/L$.

$$m_{pl}^* \simeq \frac{2\pi^2 \hbar}{\Delta\omega L^2} \qquad (3)$$

For the parameters specified in [14], $m_{pl}^*/m_o \simeq 10^{-4}$ for $n_{01} \sim 10^{16}$ m$^{-2}$ and $m_{pl}^*/m_o \simeq 3\times10^{-5}$ for $n_{01} \sim 10^{17}$ m$^{-2}$.

In the current study, we focus on the so called "low-high" plasmonic crystals where the electron density in the ungated regions is much higher than that in the gated regions. We recently demonstrated that a "low-high" grating-gate plasmonic FET architecture, with a high electron density in the ungated regions, can sustain strong resonant responses across multiple material systems [31]. In such systems, the fundamental plasmon frequency in the highly doped ungated regions $\omega_u(\pi/L_2)$ is much larger than the fundamental plasmon frequency in gated regions $\omega_g(\pi/L_1)$, and these ungated regions work as interconnecting bridges between gated plasmonic cavities [18,32]. The low-high approximation holds, when $\omega_u(\pi/L_2) \gg \omega_g(\pi/L_1)$

$$\frac{\omega_g(\pi/L_1)}{\omega_u(\pi/L_2)} = \sqrt{\pi \frac{\varepsilon+1}{\varepsilon}} \frac{\sqrt{L_2 d}}{L_1} \left(\frac{n_{01}}{n_{02}}\right)^{1/2} \ll 1 \qquad (4)$$

Usually, the effective barrier thickness, $d$, scales with the gate length $L_1$ to comply with the graduate channel approximation ($d/L_1 \sim 0.1$ to $0.25$). We also assume $(\varepsilon+1)/\varepsilon \sim 1$, and Eq. (4) reduces to

$$\frac{L_2}{L_1} \frac{n_{01}}{n_{02}} \ll \frac{4}{\pi} \qquad (5)$$

The low-high parameter region defined by Eq. (5) is presented in Fig. 3. In this region, ungated sections act as stiff, high-frequency plasmonic connectors, while the gated sections behave as low-frequency plasmonic cavities.

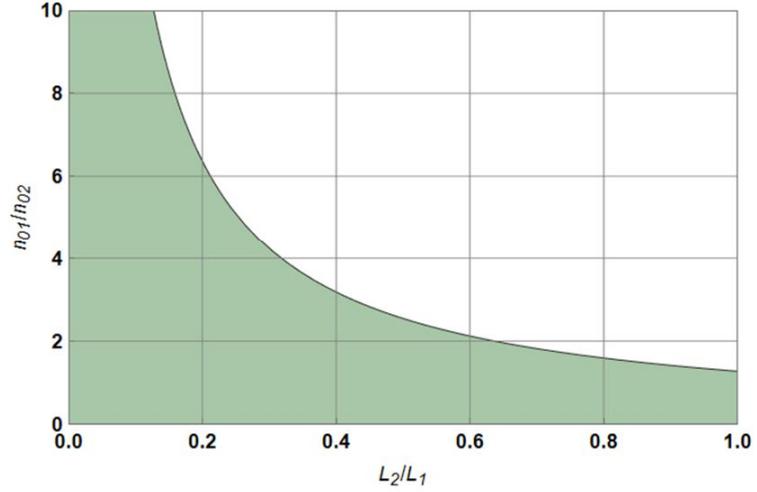

Fig. 3. The low-high parameter region in the configuration space $(L_2/L_1 , n_{01}/n_{02})$. Shaded area corresponds to the region where $L_2 n_{01}/L_1 n_{02} < 4/\pi$.

Dispersion equation for the low-high plasmonic crystal can be derived from Eq. (1) assuming strong modulation of the electron density, $\eta \ll 1$. For $a \ll 1$, in the lowest non-vanishing order on $a$ and $\eta$ we obtain

$$\cos kL = \cos\frac{\widetilde{\omega} L_1}{L} - \frac{\widetilde{\omega} \eta L_2}{2L} \sin\frac{\widetilde{\omega} L_1}{L} \qquad (6)$$

Near the center of the Brillouin zone ($kL \ll 1$) solution of this equation can be found by perturbation method

$$\omega = \omega_{\pm,n} \pm \frac{\hbar^2 k^2}{2m_{pl,n}} \qquad (7)$$

where

$$\omega_{+,n} = \frac{2\pi n v_p}{L_1} \qquad (8)$$

$$\omega_{-,n} = \frac{2\pi n v_p}{L_1}\left(1 - \frac{L_2}{L_1}\frac{n_{01}}{n_{02}}\right) \qquad (9)$$

$n = 1,2,\dots$ is the bandgap index and the plasmon effective mass is defined as

$$m_{pl,n} = \frac{n\pi\hbar L_2}{v_p L^2}\frac{n_{01}}{n_{02}} \qquad (10)$$

These results demonstrate that the band plasmon in the plasmonic crystal can be viewed as a bosonic quasiparticle similar to rotonic excitation in the superfluidity theory. This analogy leads to new possibilities for experimental and theoretical studies of this system.

Eqs. (8) and (9) represent the energies of the band plasmons at the top ($\omega_{+,n}$) and the bottom ($\omega_{-,n}$) of the $n$-th bandgap in the plasmonic crystal spectrum. The top plasmon frequency ($\omega_{+,n}$) coincides with frequency of the gated plasmon localized in the cavity under the gate. These $\omega_{+,n}$ and $\omega_{-,n}$ plasmons extend over entire plasmonic crystal as a coherent in-phase (since $k = 0$) electron density oscillations in all elementary cells and can be described as simple harmonic oscillators with respect to the electron density.

Because these band-edge plasmons extend coherently across the entire plasmonic crystal and can be treated as simple harmonic oscillators, their dynamics is especially sensitive to time-dependent modulation of the electron density. Gate voltage modulation directly affects the frequency of these oscillators, creating conditions for parametric excitation. The following section presents the analysis of parametric excitation mechanisms.

### III. PARAMETRIC EXCITATIONS IN PLASMONIC CRYSTALS

In this Section, we discuss a new technique for exciting plasmonic oscillations in the plasmonic crystal structures. The key problem in plasmonic technology is the implementation of THz or sub-THz sources, which require a finite electron drift velocity [32] controlled by a source-to-drain voltage. The problem is the resulting voltage drop across the device channel that reduces the effective gate voltage swing toward the drain decreasing plasmonic frequencies toward the drain. Voltage dividers have been proposed to alleviate this problem [9]. However, the inclusion of a gate-voltage divider substantially complicates device architecture, and to the best of our knowledge, no experimental demonstration of a plasmonic crystal FET employing such a voltage-divider scheme has been reported. Below, we analyze a new technique to excite a plasmonic crystal by periodically varying the gate voltage (using the same gate bias for all plasmonic unit cells). We show that such gate voltage pumping enables parametric resonances that might lead to instabilities in the plasmonic crystal response.

### A. Mathieu equation

The plasmonic mode anticrossings in the plasmonic crystals result in parabolic spectra at vicinity of the bandgaps, similar to roton spectrum and to a "soft-mode" spectrum near second order phase transition [33, 34]. Therefore, the plasmonic oscillations at these points could be described by the harmonic oscillator equation:

$$\frac{d^2 \delta n_\pm}{dt^2} + \frac{1}{\tau} \frac{d \delta n_\pm}{dt} + \omega_\pm^2 \delta n_\pm = 0 \tag{11}$$

where $\omega_\pm$ is defined by Eqs. (8)-(9) and $\tau$ is the momentum relaxation time.

Temporal modulation of the parameters of a harmonic oscillator may lead to the parametric resonance at certain values of the modulation frequency. In contrast to a classical parametric resonance [35], we consider a highly nonlinear parametric resonance, whose features are dramatically different from those predicted by classical theory. We demonstrate that such a nonlinear resonance could be used for frequency multiplication and effective conversion of RF modulating signal into THz signal (see also Ref. [36]).

We focus on a strongly nonlinear regime, when carrier density in the gated regions, $n_1$, is modulated by a sinusoidal gate voltage at frequency, $\delta\omega \ll \omega_\pm$ :

$$n_1(t) = n_{01} B(t) \tag{12}$$

Here, the frequency modulation function $B(t)$ is

$$B(t) = \begin{cases} 1 - A \cos \delta\omega t & if \ A \cos \delta\omega t < 1 \\ 0 & if \ A \cos \delta\omega t > 1 \end{cases} \tag{13}$$

and $A$ is the modulation amplitude. Such gate voltage pumping allows all unit cells of a plasmonic crystal to operate in unison avoiding nonuniform excitation of plasmonic unit cells by driving drain-to-source current and provides conditions for excitation of plasmons at the center of the Brillouin zone ($k = 0$). Combining Eqs. (8)-(13) we obtain the following equation describing temporal evolution of the electron density fluctuations $\delta n$ in plasma oscillations near the first bandgap in the plasmonic crystal energy spectrum in the low-high limit defined by Eq. (5):

$$\frac{d^2 \delta n}{dt^2} + \frac{1}{\tau} \frac{d \delta n}{dt} + \omega_0^2 (1 - A \cos \delta\omega t) \delta n = 0 \tag{14}$$

where

$$\omega_0 = 2\pi v_p / L_1 \tag{15}$$

is the frequency corresponding to the first ($n = 1$) gap in the plasmonic band spectrum. Eq. (14) has the form of a generalized Mathieu equation (accounting for damping). The Mathieu equation is widely used for describing parametric resonances in mechanical and electrical systems [26,27]. When $A \leq 1$, the perturbation of electron density is a simple harmonic function with frequency $\delta\omega$. According to the classical theory, it can lead to the parametric resonances at the values of $\delta\omega$ close to $\delta\omega = \frac{2\omega_0}{p}$, where $p = 1,2,\ldots$ . The width of the resonant frequency window and the instability increment rapidly decrease with increasing $p$ [35]. An additional instability decrement due to collisional damping makes this region of no practical interest. The situation changes drastically when $A > 1$. In this case, the perturbation of electron density is non-harmonic periodic function containing all frequencies that are multiples of $\delta\omega$.

Fig. 4 shows the stability phase diagram of the Mathieu equation (14) found numerically as a function of the modulation amplitude $A > 1$ and normalized damping $\gamma = 1/\omega_0\tau$. Fig. 4a is for $\delta\omega = 0.1\omega_0$, and Fig. 4b is for $\delta\omega = 0.01\omega_0$. As seen in the diagram, the unstable (pink) regions occupy significant part of the phase diagram providing for parametric resonance.

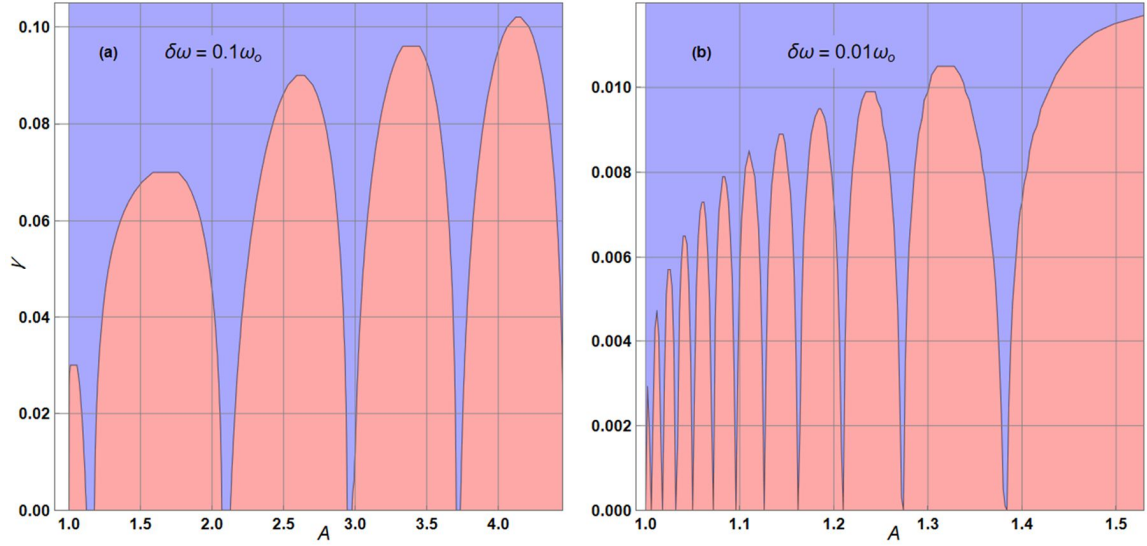

Fig. 4 The stability phase diagram of the Mathieu equation (14) in the $(\gamma, A)$ plane. $A > 1$ is the modulation amplitude, $\gamma = 1/\omega_0\tau$ is the normalized damping, and the modulation frequency $\delta\omega = 0.1\omega_0$ (a) and $\delta\omega = 0.01\omega_0$ (b). Instability regions due to the parametric resonance are marked in pink.

In the following, we use this generalized Mathieu equation for the demonstration of frequency multiplication using "low-high" AlGaN/GaN and AlGaAs/GaAs plasmonic crystals. We show that a strong nonlinear parametric resonance supports frequency multiplication and RF to THz frequency conversion. This device could overperform THz frequency multipliers relying on nonlinear properties of Schottky diodes [37].

### B. Plasmonic response to a single excitation pulse under gate voltage modulation

The analysis of the plasmonic response to a single excitation pulse under the gate voltage modulation predicts parametric plasmonic resonance in the grating-gated transistor structures. We numerically solved Mathieu equation (14) and found time-dependent plasmonic response to the initial excitation pulse $\delta n/n_{01} = 0.1$ at different gate voltage modulation frequencies $\delta\omega$ and modulation amplitudes $A$. We used the values of material and device parameters typical for AlGaAs/GaAs and AlGaN/GaN heterostructures at 77 and 300K listed in Table I.

Table I. Material and device parameters for AlGaAs/GaAs and AlGaN/GaN plasmonic crystals

| Material and Device parameters | AlGaAs/GaAs | AlGaN/GaN |
|---|---|---|
| Dielectric constant, $\varepsilon$ | 12.8 | 8.9 |
| Effective mass, $m^*$ | $0.067m_0$ | $0.24m_0$ |
| Barrier layer thickness, $d$ | 25 nm | 25 nm |
| Electron density under the gate, $n_{01}$ | $1.5 \times 10^{15}$ m$^{-2}$ | $1 \times 10^{16}$ m$^{-2}$ |
| Mobility at 77 (300) K, $\mu$ | 10 (0.7) m$^2$/V·s | 0.75 (0.2) m$^2$/V·s |
| Electron relaxation time at 77 (300)K, $\tau$ | 3.81 (0.27) ps | 1.02 (0.27) ps |
| Plasma velocity $v_p$ under the gate | $3.3 \times 10^5$ m/s | $6.1 \times 10^5$ m/s |

We present the results of our numerical simulations for both GaAs- and GaN-based structures at 77K and 300K in Fig. 5. In Figs. 5a and 5b we plotted the modulation function $B(t)$ with amplitude $A = 1.5$ and frequency $\delta\omega = 0.1\omega_0$ used in our simulations for GaAs and GaN structures, respectively. Figs. 5c and 5d show the time-dependent plasmonic responses induced in these structures by the gate modulation function $B(t)$ at 77K. Figs. 5e and 5f show the response for 300K. Here, we used the grating period $L = 1\mu$m for GaAs structures and $L = 500$ nm for GaN structures. We also assumed $L_1 = L_2 = L/2$ corresponding to the fundamental plasma frequencies $f_0 = \omega_0/2\pi$ of 745 GHz and 2.44 THz for GaAs and GaN structures, respectively. At these values of the parameters the normalized damping $\gamma$ for GaAs-based structures is equal to 0.056 at 77K and 0.80 at 300K. The normalized damping $\gamma$ for GaN-based structures is equal to 0.064 at 77K and 0.80 at 300K.

It follows from these results and the phase diagram in Fig. 4a that the plasma oscillations in both GaAs- and GaN-based structures at 77K are unstable (corresponding to the instability region predicted by the Mathieu equation) whereas at 300K the plasma oscillations are stable.

This conclusion is confirmed by direct numerical simulations of temporal evolution of the plasmonic response shown in Figs. 5c and 5d for 77K and in Figs. 5e and 5f for 300K. At 77K, the instability increment due to a parametric resonance exceeds collisional damping leading to an increasing amplitude of the plasmonic response. In this regime, the device operates as a THz oscillator. At 300K, the amplitude of the plasmonic response decreases exponentially due to increased collisional damping that exceeds the parametric resonance increment.

The demonstrated instability due to parametric resonance develops at all values of the modulation amplitude $A$ and damping $\gamma = 1/\omega_0\tau$ lying within the unstable (pink) regions on the phase diagram in Fig. 4a.

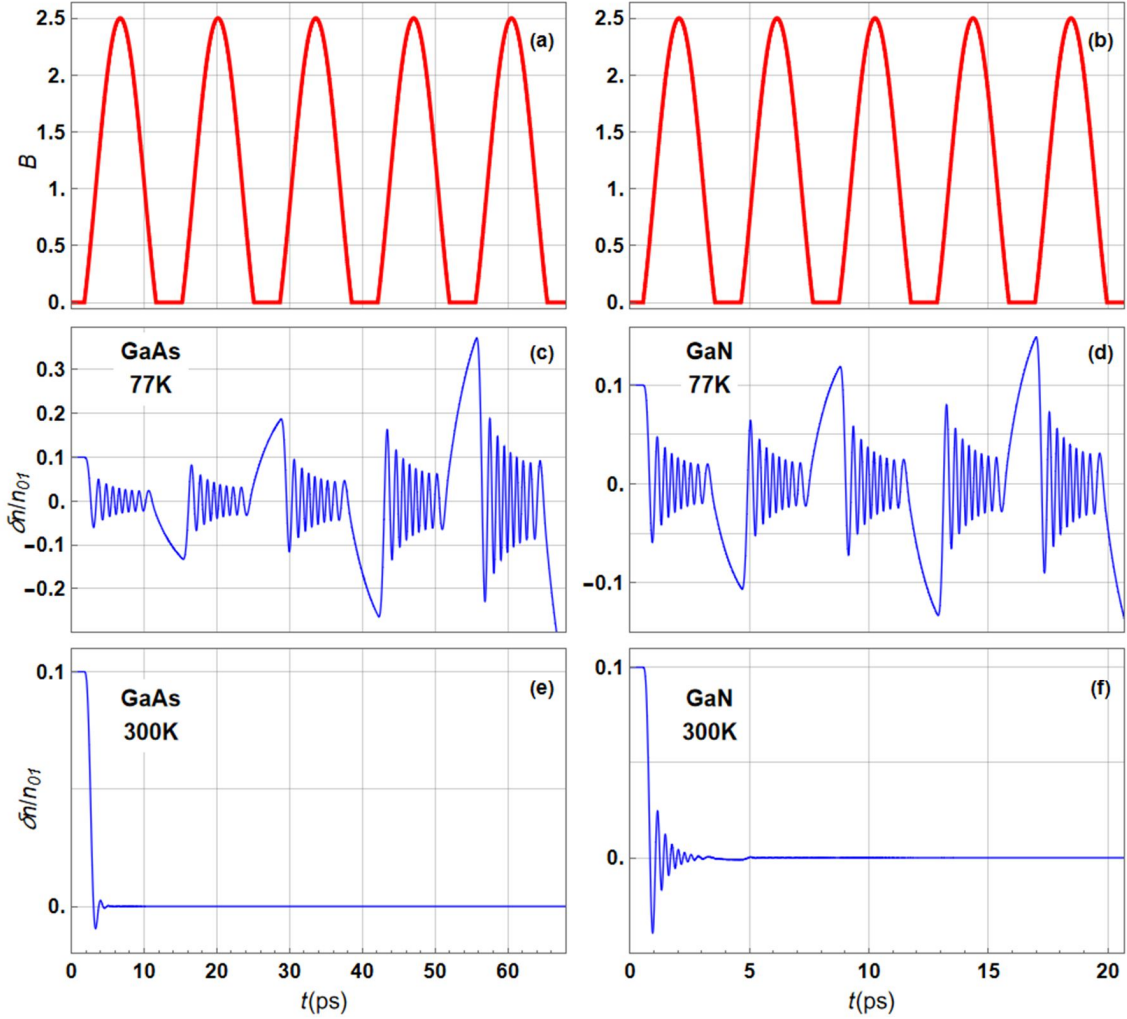

Fig. 5. Gate modulation function $B(t)$ of frequency $\delta\omega = 0.1\omega_0$ and amplitude $A = 1.5$ in the GaAs structure (a) and in the GaN structure (b). Plasmonic responses $\delta n(t)/n_{01}$ to the initial excitation pulse of $\delta n/n_{01} = 0.1$ under the gate modulation in the GaAs structure with grating period $L = 1$ μm at 77K (c), 300K (e) and in the GaN structure with grating period $L = 500$ nm at 77K (d), 300K (f). Here, $f_0 = \omega_0/2\pi$ is the fundamental plasma frequency under the gates: $f_0 = 0.745$ (2.44) THz in GaAs (GaN) structures. Other material and device parameters are listed in Table I.

At very low temperatures (~4K) the mobilities in the GaAs-based heterostructures are very high, and the instability due to parametric resonance can occur even at lower modulation frequencies. As an illustration, Fig. 6 presents plasmonic response of the GaAs structure with the same set of parameters as in Fig. 5 but for the mobility value of $\mu = 100$ m$^2$/Vs.

Fig. 6a shows the gate modulation function $B(t)$ with the amplitude $A = 1.5$ and the modulation frequency $\delta\omega = 0.01\omega_0$.

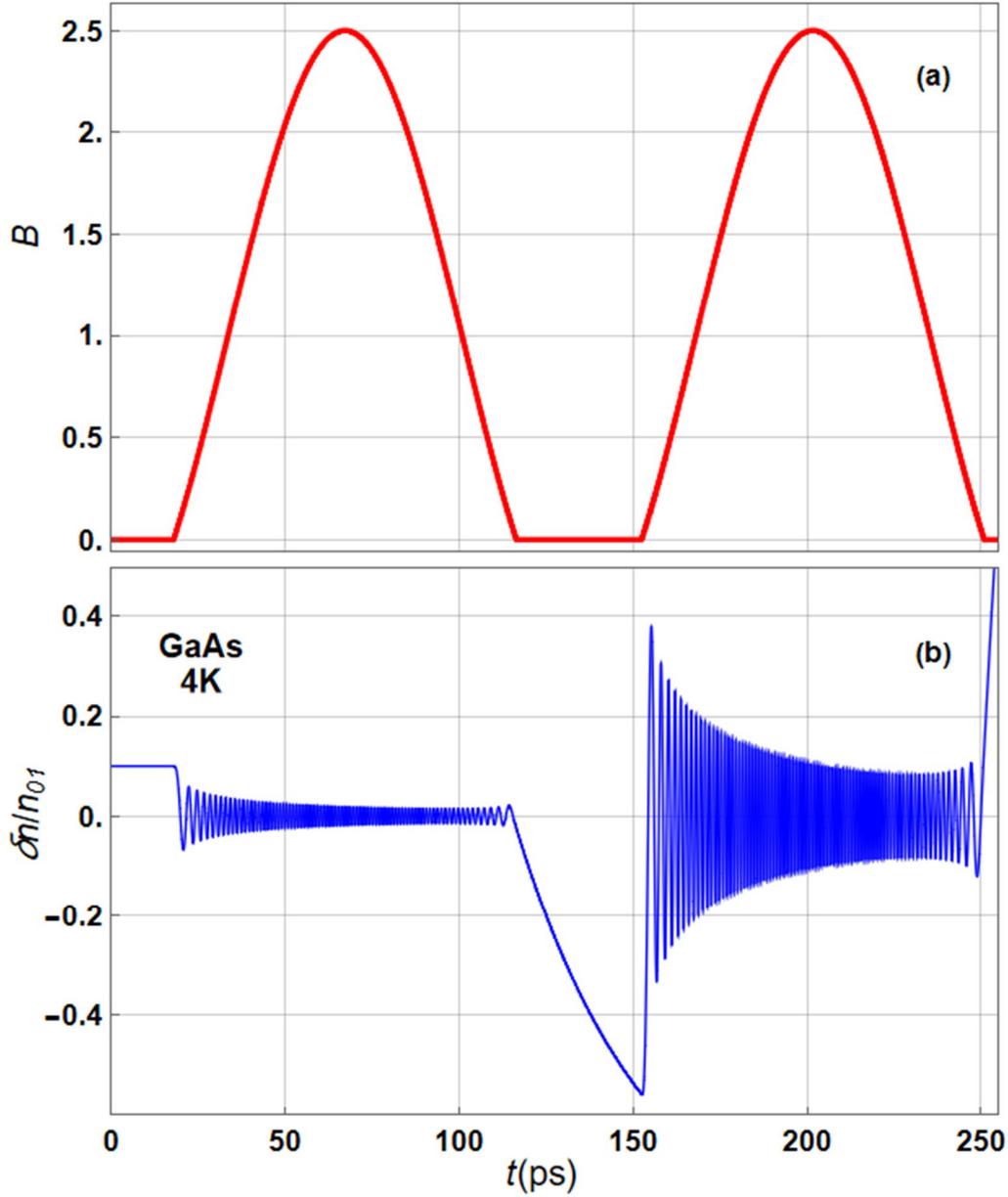

Fig. 6. Modulation function $B(t)$ for $\delta\omega = 0.01\omega_0$ and amplitude $A = 1.5$ (a); Plasmonic response $\delta n(t)/n_{01}$ to the initial excitation pulse of $\delta n/n_{01} = 0.1$ under the gate modulation $B(t)$ in GaAs structure with grating period $L = 1$ μm and mobility $\mu = 100$ m²/V s at 4K (b). Here, $f_0 = \omega_0/2\pi = 0.745$ THz is the fundamental plasma frequency under the gates. Table I lists other material and device parameters

In this case, the normalized damping $\gamma = 0.0056$, and the plasmonic response falls into the instability region shown in Fig. 4b. The plasmonic response in Fig. 6b reveals the instability due to parametric resonance demonstrating the conversion of the ~10 GHz modulation signal into the ~1 THz plasmonic signal.

Damping increases with temperature, and, as shown in Figs. 5e and 5f, there is no instability at elevated temperatures. In order to provide THz generation in these structures at room temperatures one should move to larger plasma frequencies $f_0$ and/or larger modulation amplitudes, see Fig. 4.

Figs. 7a and 7b show the gate modulation function $B(t)$ at 300K with an amplitude $A = 3.4$ and $\delta\omega = 0.1\omega_0$ in GaAs and GaN systems (Table I lists the parameters used in the calculation). Figs. 7c and 7d show the resulting response to the gate modulation $B(t)$. In this calculation, we also decreased the grating period down to $L = 100$nm for the GaAs structures and $L = 200$nm for the GaN structures. The smaller grating period results in larger fundamental plasma frequencies $f_0$ in Eq. (15): $f_0 = 7.45$ THz in the GaAs structure and $f_0 = 6.10$ THz in the GaN structure. These frequencies are still lower than the frequencies of the optical phonons in GaAs ($\geqslant 8.0$ THz) and in GaN ($\geqslant 16$ THz) so that the plasma oscillations at these frequencies do not experience an additional damping due to scattering on optical phonons. The results on Figs. 7c and 7d demonstrate the instability which opens the possibility to use parametric resonance in the plasmonic crystal structures for THz generation at room temperature.

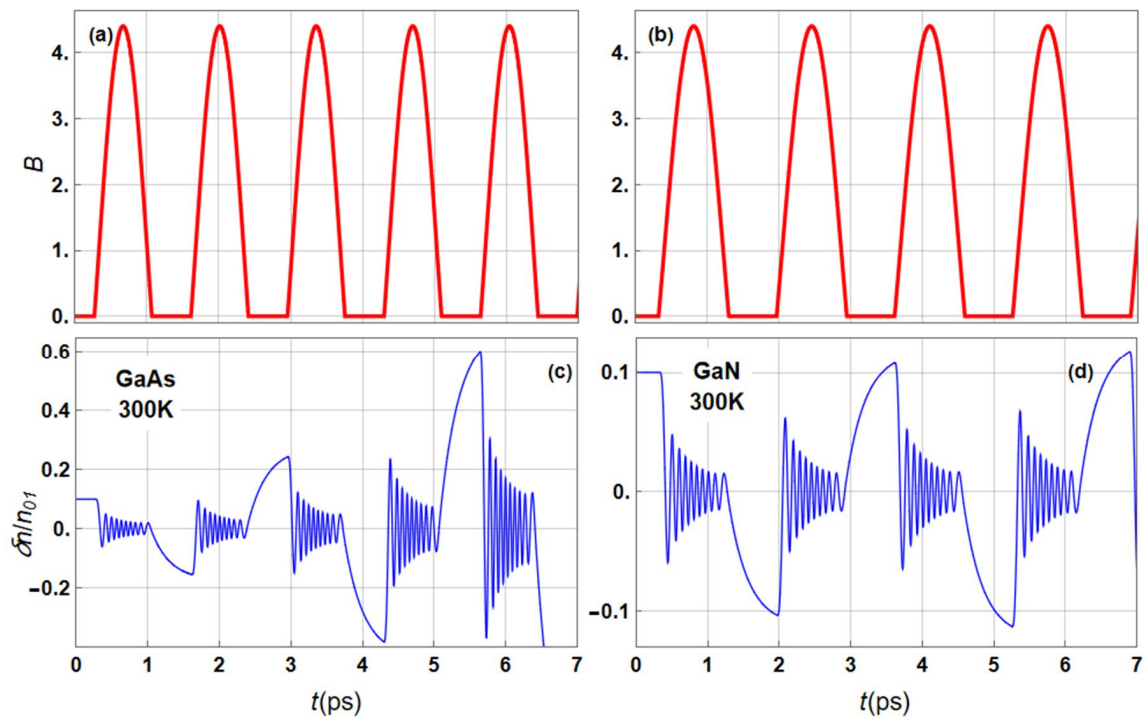

Fig. 7. Gate modulation function $B(t)$ of frequency $\delta\omega = 0.1\omega_0$ and amplitude $A = 3.4$ in the GaAs structure (a) and in the GaN structure (b). Plasmonic responses $\delta n(t)/n_{01}$ to the initial excitation pulse of $\delta n/n_{01} = 0.1$ under the gate modulation in the GaAs structure with grating period $L = 100$ nm (c) and in the GaN structure with grating period $L = 200$ nm (d) at 300K. Here, $f_0 = \omega_0/2\pi$ is the fundamental plasma frequency under the gates: $f_0 = 7.45$ THz in GaAs structures and $f_0 = 6.10$ THz in GaN structures. Other material and device parameters are listed in Table I.

The results shown in this section demonstrate the parametric instability of THz plasma oscillations in plasmonic crystals under the gate voltage modulation. Since equations of the model were derived for small fluctuations of the electron density the developed theory can predict temporal evolution of the plasmonic response only during limited time interval after the initial excitation when $\delta n/n_{01} \ll 1$. Description of subsequent saturation of the response and final stationary radiating state requires inclusion of higher order nonlinearities and radiative losses into the theory. This problem is beyond the scope of the current paper and will be considered elsewhere.

## IV. CONCLUDING REMARKS

We present a theory describing plasma excitations with parabolic dispersion in periodic gated-ungated field-effect structures based on III-N and III-V material platforms (plasmonic crystals). We call such plasma excitations rotonic plasmons due to the similarity of the energy spectrum of these bosonic quasiparticles with the spectrum of rotons in the theory of superfluidity. These excitations emerge because of a strong coupling between alternating gated and ungated plasmonic regions supporting plasmon's band diagram. Near the band edges, rotonic plasmons have parabolic dispersion characterized by an effective plasmonic mass. This behavior is in contrast to the linear or square-root dispersion in purely gated or ungated structures.

A periodic gate-voltage modulation modulates the frequency of rotonic plasmons. Such a behavior could be described by solutions of the generalized Mathieu equation, which accounts not only for periodic plasmon frequency modulation, but also for damping.

By solving such generalized Mathieu equation under periodic gate-voltage modulation, we demonstrated that, depending on damping, rotonic plasmons can support both resonant and non-resonant parametric excitations. Above a critical excitation level (sufficient to overcome damping) these excitations lead to efficient parametric instabilities in plasmonic crystals. This is achieved without current-driven excitation that is non-uniform and lossy.

We propose using the predicted parametric resonances to generate plasma oscillations at THz frequencies. This approach minimizes spatial nonuniformities (unavoidable when using current-driven pumping) and enables THz power generation.

Our simulations show that this mechanism enables parametric THz generation and RF-to-THz frequency multiplication in a wide temperature range including cryogenic and room temperatures in III-N, and III-V HEMT plasmonic crystals.

Simulations confirm that gate-voltage modulation in these material systems can drive self-sustained oscillations. The frequency of these oscillations for GaN and GaAs grating-gated structures with the period shorter than approximately 500 nm is in THz range of frequency. These oscillations are especially pronounced in the low–high carrier density regime. This behavior could be observed across various III-N and III-V heterostructures and graphene- and silicon-based implementations.

Our results position rotonic plasmons as key enablers for compact, tunable, and high-efficiency RF and THz sources suitable for emerging 6G communication and high-resolution sensing systems. The combination of parabolic dispersion, nonlinear dynamics, and CMOS compatibility makes plasmonic crystals a versatile platform for bridging the THz gap.

## APPENDIX

In this Appendix, for the reference purposes, we derive dispersion equation, Eq. (1), for the plasmonic crystal formed in the 2D electron channel of the transistor nanostructure with alternating gated and ungated sections shown in Fig. 1. In the hydrodynamic approximation, plasmon dynamics in the 2D electron channel with equilibrium electron density $n_0$ is described by the Euler equation and the equation of continuity for small fluctuations of electron density $\delta n$ and hydrodynamic velocity $\delta v$:

$$\begin{cases} \frac{\partial \delta v}{\partial t} = \frac{e}{m^*} \frac{\partial \delta \varphi}{\partial x} - \frac{\delta v}{\tau} \\ \frac{\partial \delta n}{\partial t} + n_0 \frac{\partial \delta v}{\partial x} = 0 \end{cases} \qquad (A1)$$

Here, $\delta \varphi$ is electric potential in the channel due to fluctuation of the charge density $\delta \rho = -e \delta n$. All other notations below are defined in the main text.

We look for solutions of Eq. (A1) in the form $\delta n = \delta n_{q\omega} e^{-iqx+i\omega t}$ and $\delta v = \delta v_{q\omega} e^{-iqx+i\omega t}$. In the quasistatic approximation and at $qd \ll 1$ electric potential $\delta \varphi_{q\omega}$ and fluctuation of the electron charge density $\delta \rho_{q\omega}$ are connected by the local capacitance $C$: $\delta \rho_{q\omega} = C \delta \varphi_{q\omega}$. In the gated channels, this capacitance is the local gate capacitance $C_g = \frac{\varepsilon \varepsilon_0}{d}$, and

$$\delta \rho_{q\omega} = \frac{\varepsilon \varepsilon_0 \delta \varphi_{q\omega}}{d} \qquad \text{(gated)} \qquad (A2)$$

In the ungated channels, the local capacitance is $C_u = (\varepsilon + 1)\varepsilon_0 q$ [8, 38]. This expression for $C_u$ accounts for the boundary between the dielectric layer and free space ($\varepsilon = 1$) leading to the replacement of the dielectric permittivity $\varepsilon$ by the average permittivity $(\varepsilon + 1)/2$. In this case,

$$\delta \rho_{q\omega} = (\varepsilon + 1)\varepsilon_0 q \delta \varphi_{q\omega} \qquad \text{(ungated)} \qquad (A3)$$

Substituting Eq. (A2) or (A3) into Eq. (1) we obtain a homogeneous system of linear algebraic equations for unknown $\delta \varphi_{q\omega}$ and $\delta v_{q\omega}$. In the limit of small damping ($\omega_{pl} \tau \gg$ 1), general solution of this system for the local current density $\delta j_{q\omega} = -en_0 \delta v_{q\omega}$ and local potential $\delta \varphi_{q\omega}$ in gated ($-L_1 \leq x \leq 0$) and ungated ($0 \leq x < L_2$) sections of the 2D channel in Fig. 1 is

$$
\begin{cases}
\delta \varphi_g(x) = A_1 e^{-iq_1 x} + A_2 e^{iq_1 x} \\
\delta j_g(x) = \frac{\omega C_g}{q_1} A_1 e^{-iq_1 x} - \frac{\omega C_g}{q_1} A_2 e^{iq_1 x}
\end{cases}
\quad -L_1 \leq x \leq 0 \quad \text{(gated)} \quad \text{(A4)}
$$

$$
\begin{cases}
\delta \varphi_u(x) = A_3 e^{-iq_2 x} + A_4 e^{iq_2 x} \\
\delta j_u(x) = \frac{\omega C_u}{q_2} A_3 e^{-iq_2 x} - \frac{\omega C_u}{q_2} A_4 e^{iq_2 x}
\end{cases}
\quad 0 \leq x \leq L_2 \quad \text{(ungated)} \quad \text{(A5)}
$$

where the plasmon wave vectors $q_1$ and $q_2$ at given frequency $\omega$ are determined by Eq. (2) for gated and ungated sections, respectively. Unknown coefficients $A_1, \ldots A_4$ are determined by the boundary conditions at the boundaries between gated and ungated sections.

In our calculations we used ballistic boundary conditions. These conditions apply if the width of the transition region between gated and ungated sections of the channel is less than the electron-electron scattering length (see detailed discussion in Ref. [28]). In the opposite limit the hydrodynamic boundary conditions are more appropriate. Using the hydrodynamic boundary conditions changes some numerical factors in the equations but does not qualitatively change the plasmon dispersion law [8]. Ballistic boundary conditions imply continuity of the electron current $\delta j$ and electric potential $\delta \varphi$ at the boundaries between gated and ungated sections positioned at $x = 0$ and $x = L_2$ in Fig. 1:

$$\delta \varphi_g(0) = \delta \varphi_u(0); \;\; \delta j_g(0) = \delta j_u(0); \; \delta \varphi_g(L_2) = \delta \varphi_u(L_2); \;\; \delta j_g(L_2) = \delta j_u(L_2); \quad \text{(A6)}$$

These equations should be supplemented by the Bloch boundary condition for the periodic system

$$\delta \varphi_g(L_2) = \delta \varphi_g(-L_1) e^{-ikL}; \;\; \delta j_g(L_2) = \delta j_g(-L_1) e^{-ikL}; \quad \text{(A7)}$$

where $k$ is the Bloch wave vector and $L$ is the period of the structure.

Combining Eqs. (A4) – (A7) we arrive to the following system of linear algebraic equations for coefficients $A_1, \ldots A_4$:

$$A_1 + A_2 - A_3 - A_4 = 0$$

$$A_1 - A_2 - \frac{q_1 C_u}{q_2 C_g} A_3 + \frac{q_1 C_u}{q_2 C_g} A_4 = 0$$

$$e^{-iq_1 L_1} e^{ikL} A_1 + e^{-iq_1 L_1} e^{-ikL} A_2 - e^{-iq_2 L_2} A_3 - e^{iq_2 L_2} A_4 = 0 \quad \text{(A8)}$$

$$e^{-iq_1 L_1} e^{ikL} A_1 - e^{-iq_1 L_1} e^{-ikL} A_2 - \frac{q_1 C_u}{q_2 C_g} e^{-iq_2 L_2} A_3 + \frac{q_1 C_u}{q_2 C_g} e^{iq_2 L_2} A_4 = 0$$

Determinantal equation for the system (A8) yields Eq. (1) in the main text.